\documentclass[showpacs,preprintnumbers,
amsmath,amssymb,groupedaddress,superscriptaddress]{revtex4}
\usepackage{graphicx}
\usepackage{dcolumn}
\usepackage{bm}

\begin{document}
%\draft 
\title{Running sums for $2\nu\beta\beta$-decay matrix elements within the QRPA with account for deformation}

\author{Dongliang Fang}
\affiliation{Institute f\"{u}r Theoretische Physik der Universit\"{a}t
T\"{u}bingen, D-72076 T\"{u}bingen, Germany}
\author{Amand Faessler}
%\email{amand.faessler@uni-tuebingen.de}
\affiliation{Institute f\"{u}r Theoretische Physik der Universit\"{a}t
T\"{u}bingen, D-72076 T\"{u}bingen, Germany}
\author{Vadim Rodin}
\email{vadim.rodin@uni-tuebingen.de}
\affiliation{Institute f\"{u}r Theoretische Physik der Universit\"{a}t
T\"{u}bingen, D-72076 T\"{u}bingen, Germany}
\author{Mohamed Saleh Yousef}
\affiliation{Department of Physics, University of Cairo,
%Orman, Ghiza
Cairo, Egypt}
\author{Fedor \v Simkovic}
%\email{fedor.simkovic@fmph.uniba.sk}
\altaffiliation{On  leave of absence from Department of Nuclear
Physics, Comenius University, Mlynsk\'a dolina F1, SK--842 15
Bratislava, Slovakia}
\affiliation{Institute f\"{u}r Theoretische Physik der Universit\"{a}t
T\"{u}bingen, D-72076 T\"{u}bingen, Germany}

\date{\today}

\begin{abstract}
The $2\nu\beta\beta$-decay running sums for $^{76}$Ge and $^{150}$Nd nuclei are calculated within a QRPA approach with account for deformation. A realistic nucleon-nucleon residual interaction based on the Brueckner G matrix (for the Bonn CD force) is used. 
%%%%%%%%%%%%%%%%%%%%%%%%%%%%
The influence of different model parameters on the functional behavior of the running sums is studied.
%%%%%%%%%%%%%%%%%%%%%%%%%%%%
It is found that the parameter $g_{pp}$ renormalizing the G matrix in the QRPA particle-particle channel is responsible for a qualitative change in behavior of the running sums at higher excitation energies. For realistic values of $g_{pp}$ a significant negative contribution to the total $2\nu\beta\beta$-decay matrix element is found to come from the energy region of the giant Gamow-Teller resonance. 
%%%%%%%%%%%%%
This behavior agrees with the results of other authors.
%%%%%%%%
\end{abstract}



\date{\today}

\maketitle

%\section{\bf Introduction.} 
Neutrinoless double beta ($0\nu\beta\beta$) decay, if observed, can allow one to extract the absolute neutrino mass from the decay rate provided that the corresponding nuclear matrix elements can be calculated reliably (see, e.g.,~\cite{fae98,Rod05}). But up to now there has been no complete agreement among different many body methods calculating these matrix elements. Thus, further tests of the reliability of the calculated nuclear wave functions are needed.
A probe that is of particular relevance is two-neutrino double beta ($2\nu\beta\beta$) decay, as it connects
the same initial and final nuclear ground states that $0\nu\beta\beta$ decay does. The corresponding nuclear transition amplitudes
$M^{2\nu}_{exp}$ have been determined accurately in several nuclei of interest (see, e.g.~\cite{barab} and references therein).

The total $2\nu\beta\beta$-decay matrix element $M^{2\nu}$     %$2\nu\beta\beta$ transition matrix element 
is dominated by the double Gamow-Teller (GT) transition matrix element $M^{2\nu}_{GT}$. The latter is
determined by the amplitudes of the single GT transitions connecting the initial and final ground states with all $1^+$ states in the intermediate nucleus. 
The GT strength distributions in the first and second legs of the decay provide a more detailed, differential test of the calculations (while $M^{2\nu}$ is an integral quantity). It is customary to represent the partial contribution of all $1^+$ states with the excitation energies $E_x\le \omega$ to the total $M^{2\nu}$ in terms of the so-called $2\nu\beta\beta$ running sums $M^{2\nu}(\omega)$.
%which allow more detailed comparison of different calculations of $M^{2\nu}$. For instance, it can be 
The running sum allows one to test whether so-called single-state dominance (SSD) is realized in calculations. The SSD hypothesis states that the single transition via the lowest $1^+$ state in the intermediate nucleus can account for the entire $M^{2\nu}$.

For the three double-$\beta$-decay systems $A=100,\ 116,$ and 128, in which the ground state of the intermediate nucleus has $J^\pi=1^+$,  the single $\beta$ decay in the second leg and the electron capture for the first leg provide additional tests of the quality of the theoretical calculations~\cite{fae07}.
Recently, the GT strengths relevant for $2\nu\beta\beta$ decay have been studied experimentally by means of different charge-exchange reactions~\cite{madey,aki97,frekers,yako09}, 
thus providing important detailed information on the GT strength distribution.
These measurements have shown that in some nuclei the SSD is indeed realized. This means that the total matrix element of $2\nu\beta\beta$ decay for those nuclei can be reconstructed fully from the measurement results only. For all the nuclei it has been obtained that the dominating contribution to $M^{2\nu}$ comes essentially from the low-lying GT strength and in such a case charge-exchange reaction studies allow the provision of an upper limit on $M^{2\nu}$.

The $2\nu\beta\beta$ running sums calculated for $^{48}$Ca within the nuclear shell model~\cite{Zhao90,Hor07,Nak96,Caur04} consistently show a substantial overshoot  
of the total $M^{2\nu}_{GT}$ at excitation energies about 7 MeV, which then is compensated by a negative contribution coming from the region of the giant GT resonance (GTR). The total $M^{2\nu}_{GT}=0.054$ MeV$^{-1}$ calculated in Ref.~\cite{Hor07} is in excellent agreement with the corresponding experimental value, $M^{2\nu}_{exp}=0.05\pm 0.01$ MeV$^{-1}$~\cite{barab}.
The same feature was also observed in the QRPA calculations for $^{76}$Ge and $^{100}$Mo~\cite{Rod05}. In contrast, recent calculations of $2\nu\beta\beta$ running sums by Moreno et al.~\cite{Mor08}, systematically  performed within the QRPA with a schematic residual interaction for many nuclei, do not reveal any destructive contribution coming from the GTR region. 
%%%%%%%%
Puzzled by this difference, we study in this Brief Report the influence of different QRPA model parameters on the functional behavior of the running sums by utilizing a QRPA approach developed in Ref.~\cite{Sal09}. The approach allows one to use, for deformed nuclei, a realistic effective interaction based on the Brueckner $G$ matrix derived from the nucleon-nucleon Bonn CD force.
%%%%%%%%
We present an analysis of the $2\nu\beta\beta$ running sums calculated for $^{76}$Ge and $^{150}$Nd within the QRPA approach of Ref.~\cite{Sal09}. 
We find that the parameter $g_{pp}$ renormalizing the strength of the G matrix interaction in the particle-particle channel is responsible for the change in sign of the contribution to the total $M^{2\nu}_{GT}$ coming from high-lying states.
%%%%%%%%
Thus, different choice of the $g_{pp}$ parameter in different QRPA calculations may lead to a different, constructive or destructive, contribution to the total $2\nu\beta\beta$-decay matrix element coming from high-lying $1^+$ states.
Therefore, a realistic value of $g_{pp}$ should be chosen thoroughly within a QRPA approach before a conclusion regarding the validity of the SSD is drawn. For example, quenching of the axial-vector constant $g_A$ can play an important role in fixing 
$g_{pp}$. For realistic values of $g_{pp}$ in the present calculation (neglecting the quenching effect), a significant negative contribution to the total $2\nu\beta\beta$-decay matrix element is found to come from the energy region of the GTR. The values of $g_{pp}$ adopted in Ref.~\cite{Mor08} are apparently smaller than in the present calculations, and this may be a reason for the aforementioned differences in the functional behavior of the calculated running sums.

%\section {\bf Basic formulas.}
A detailed description of the QRPA approach employed here as well as the choice of the model parameters is given in Ref.~\cite{Sal09}. Here, we give only the formulas that are important for the present calculation of $2\nu\beta\beta$ running sums $M^{2\nu}(\omega)$.

The matrix element $M^{2\nu}_{GT}$ within the QRPA is given in the intrinsic system by the following expression~\cite{Sal09}:
\begin{equation}
M^{2\nu}_{GT}= \sum_{K=0,\pm 1} \sum_{{m_i m_f}}
\frac{\langle 0^+_f| \bar\beta^-_{K} | K^+,m_f\rangle\langle K^+,m_f|K^+,m_i\rangle
\langle K^+,m_i| \beta^-_K | 0^+_i\rangle}{\bar\omega_{K,m_im_f}}.
\label{M2nu}
\end{equation}
Here, $K$ is the projection of the total angular momentum onto the nuclear symmetry axis, characterizing nuclear excitations in the intrinsic system, $\langle 0^+_f| \bar\beta^-_{K} | K^+,m_f\rangle$ and $\langle K^+,m_i| \beta^-_K | 0^+_i\rangle$ are the transition matrix elements of the GT operator $\beta^-_K$ (an explicit representation of these matrix elements as well as of the overlap $\langle K^+,m_f|K^+,m_i\rangle$  is given in Ref.~\cite{Sal09}). As in Ref.~\cite{Sal09}, we shift the whole calculated QRPA energy spectrum in such a way as to have the first calculated $1^+$ state exactly at the corresponding experimental energy. In this case the energy denominator in Eq.~(\ref{M2nu}) acquires the form ${\bar\omega_{K,m_im_f}}=(\omega_{K,m_f} - \omega_{K,1_f} + \omega_{K,m_i}  - \omega_{K,1_i})/2+\bar\omega_{1^+_1}$, with $\bar\omega_{1^+_1}$ being the experimental energy of the first $1^+$ state relative to the mean ground-state energy of the initial and final nuclei, and $(E_{0_i}+E_{0_f})/2$ and $\omega_{K,m_f}(\omega_{K,m_i})$ being the calculated QRPA energies relative to the initial (final) nucleus. All the calculated running sums in this work are represented in terms of $\bar\omega$. 

The $2\nu\beta\beta$ running sum $M^{2\nu}_{GT}(\omega)$  is defined as the sum, Eq.~(\ref{M2nu}),
truncated by the condition $\bar\omega_{K,m_im_f}\le \omega$.

%\section {\bf Results.}
In the present work we adopt the model parameters of Ref.~\cite{Sal09}. 
A realistic residual interaction based on the Brueckner $G$ matrix derived from the nucleon-nucleon Bonn CD force is used in the calculation, along with a schematic separable one. 
The residual interaction is renormalized in both the particle-hole and the particle-particle channels by means of multiplicative factors $g_{ph}$ and $g_{pp}$, respectively 
(corresponding strength constants in the case of the schematic force are $\chi$ and $\kappa$).
The particle-hole renormalization factor $g_{ph}$
is determined by fitting the experimental position of the GTR in the intermediate nucleus.
For both $^{76}$Ge and $^{150}$Nd the corresponding values $g_{ph}^0=1.15$ for the realistic force and $\chi^0=3.73/A^{0.7}$~MeV for the phenomenological separable force were found in Ref.~\cite{Sal09}. 
The particle-particle renormalization factor $g_{pp}$ is chosen so as to fit
the experimental value $M_{GT}^{2\nu}$ for a given nucleus. 
The corresponding fitted values of $g_{pp}^0$ and $\kappa^0$ also found in Ref.~\cite{Sal09}, along with the choices of the deformation parameter $\beta_2$, are listed in Table~\ref{tab1}.

\begin{table}[h]
%\centering
\caption{Values of the deformation parameter $\beta_2$ for initial (final) nuclei adopted in the calculations along
with the fitted values of the p-p strength parameters $g_{pp}^0$ (for the realistic Bonn-CD force) and $\kappa^0$ (for a phenomenological separable force).
%The p-h strength parameters $g_{ph}=1.15$ and $\chi=3.73/A^{0.7}$ MeV are fixed as explained in the text.
}
\begin{tabular}{|r|l|c|c|c|c|}
	\hline	
Nucleus    &\ \ \ \  $\beta_2$ &                $g_{pp}^0$  & $\kappa^0$ (MeV) \\ 
	\hline
$^{76}$Ge ($^{76}$Se) & 0.0\ \ (0.0)              & 0.94  & 0.087  \\
                       & 0.10 (0.16)              & 0.99  & 0.091  \\
\hline
$^{150}$Nd ($^{150}$Sm) & 0.0\ \  (0.0)        & 1.11 & 0.051   \\
Def. I\                   & 0.37 (0.23)               & 0.78  & 0.033 \\
Def. II                   & 0.24 (0.21)               & 1.35  & 0.053 \\
\hline 
\end{tabular}
\label{tab1}
\end{table}

\begin{figure}[htb]
%\centerline{\includegraphics[width=0.7\textwidth]{gesp.eps}}
\includegraphics[scale=0.35]{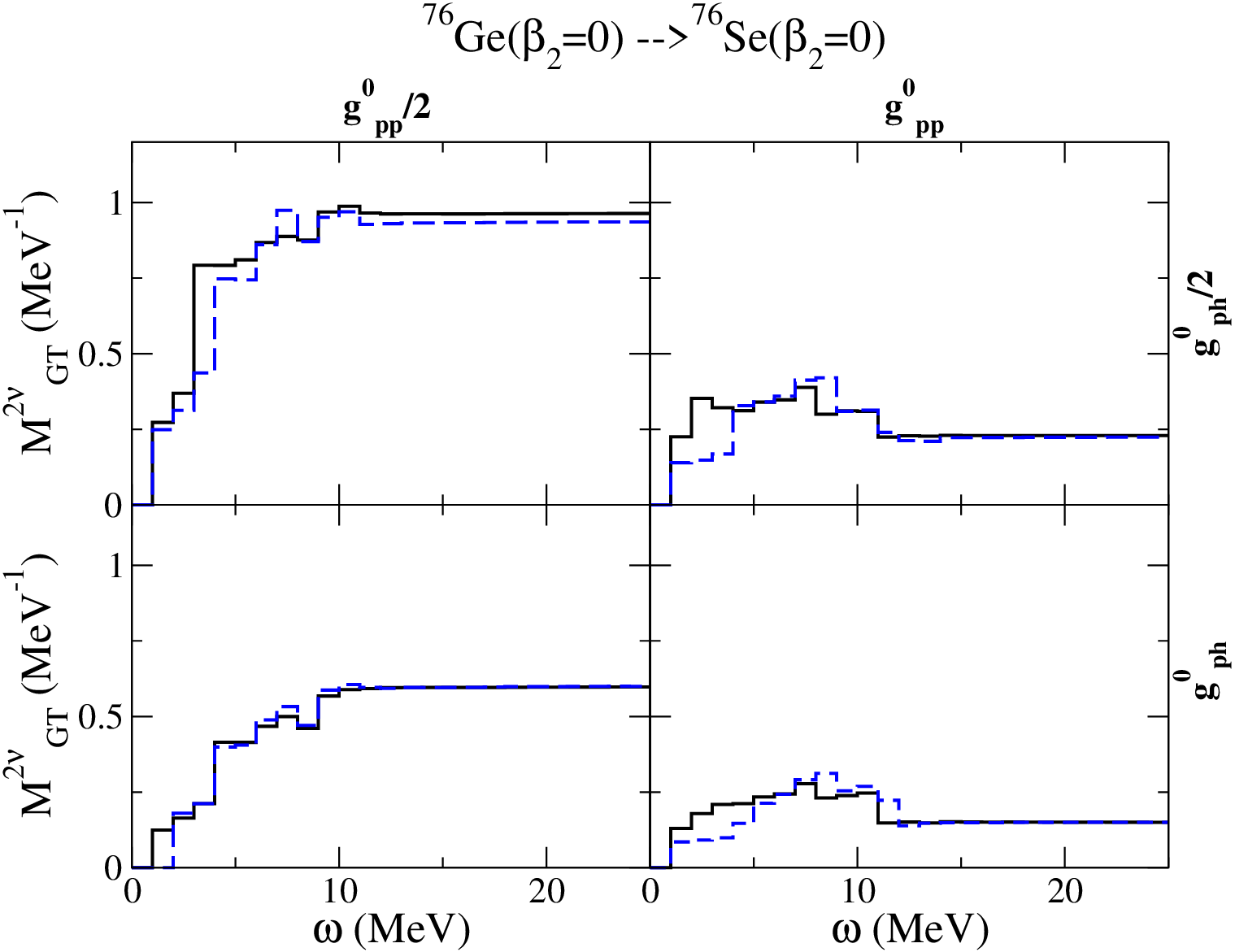}~~~~~\includegraphics[scale=0.35]{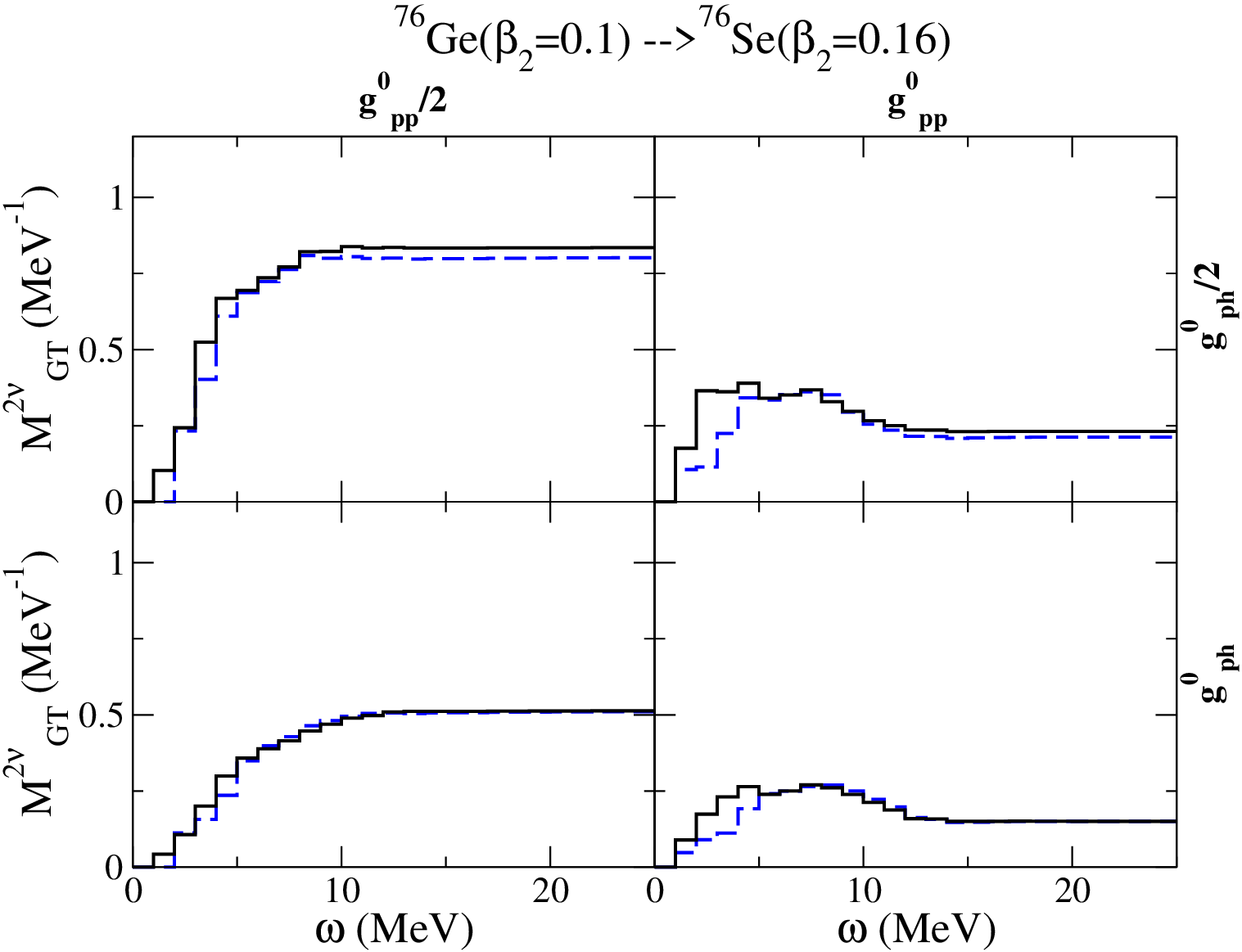}
\caption{(Color online) Running sums $M_{GT}^{2\nu}(\omega)$ for $^{76}$Ge$\to^{76}$Se $2\nu\beta\beta$ decay, 
for different choices of deformation and strength parameters.
% $g_{ph}$ and $g_{pp}$
The solid black line represents the QRPA results obtained with a realistic residual interaction ($G$ matrix from the Bonn CD force), while the dashed (blue) line corresponds to the calculation with a schematic separable force. The left and right panels correspond to different deformations of the initial and final nuclei in the calculation. Each panel consists of four sub-panels forming $2\times 2$ matrix, whose columns are marked by 
the strength of $g_{pp}$ (or $\kappa$) taken in the calculation 
[the full or a half of the fitted value $g_{pp}^0$ ($\kappa^0$) given in Table~\protect\ref{tab1}] and whose rows are labeled by the strength of $g_{ph}$ (or $\chi$) in the calculation.}
\label{ge}
\end{figure}

\begin{figure}[htb]
\includegraphics[scale=0.35]{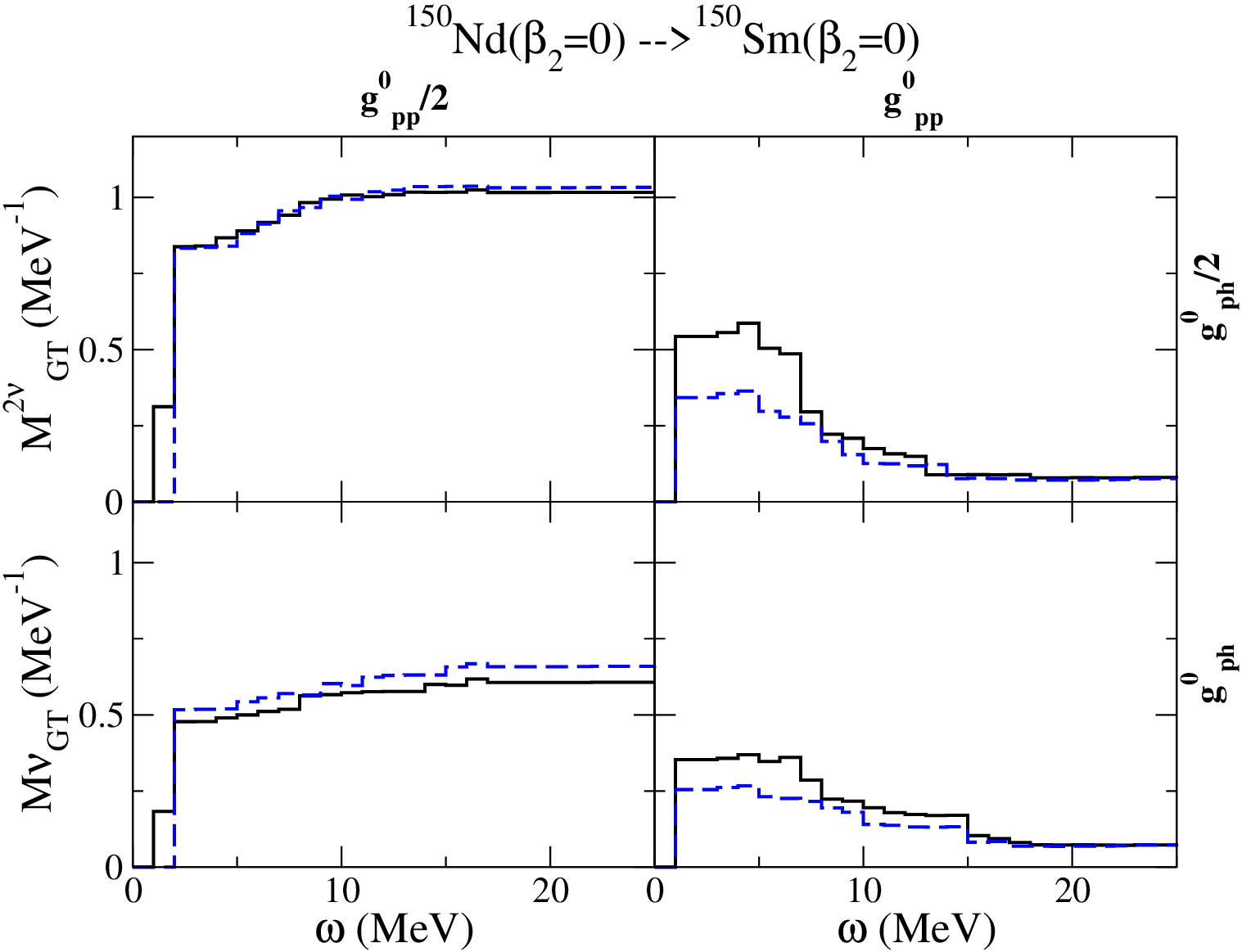}~~~~~\includegraphics[scale=0.35]{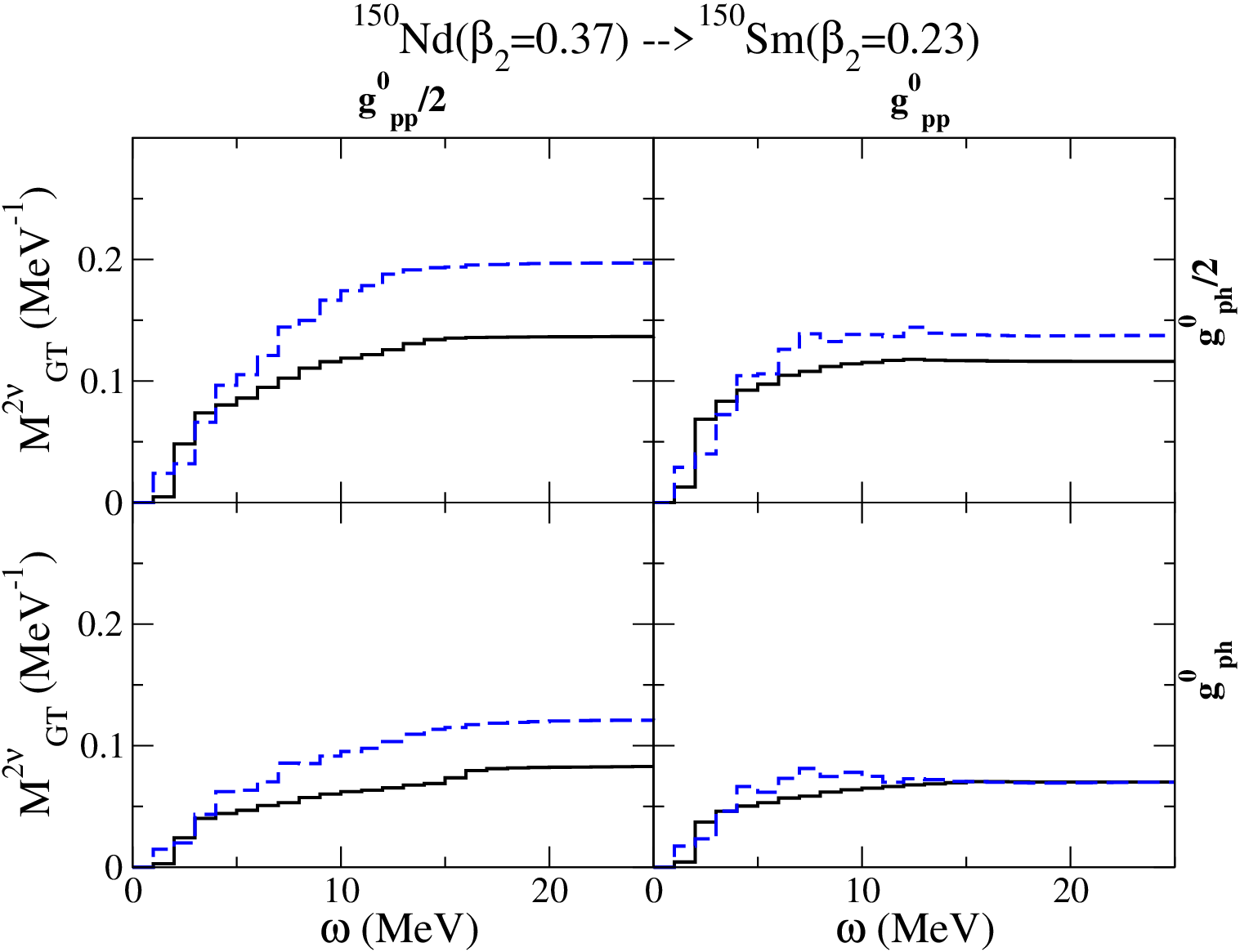}

\

\includegraphics[scale=0.35]{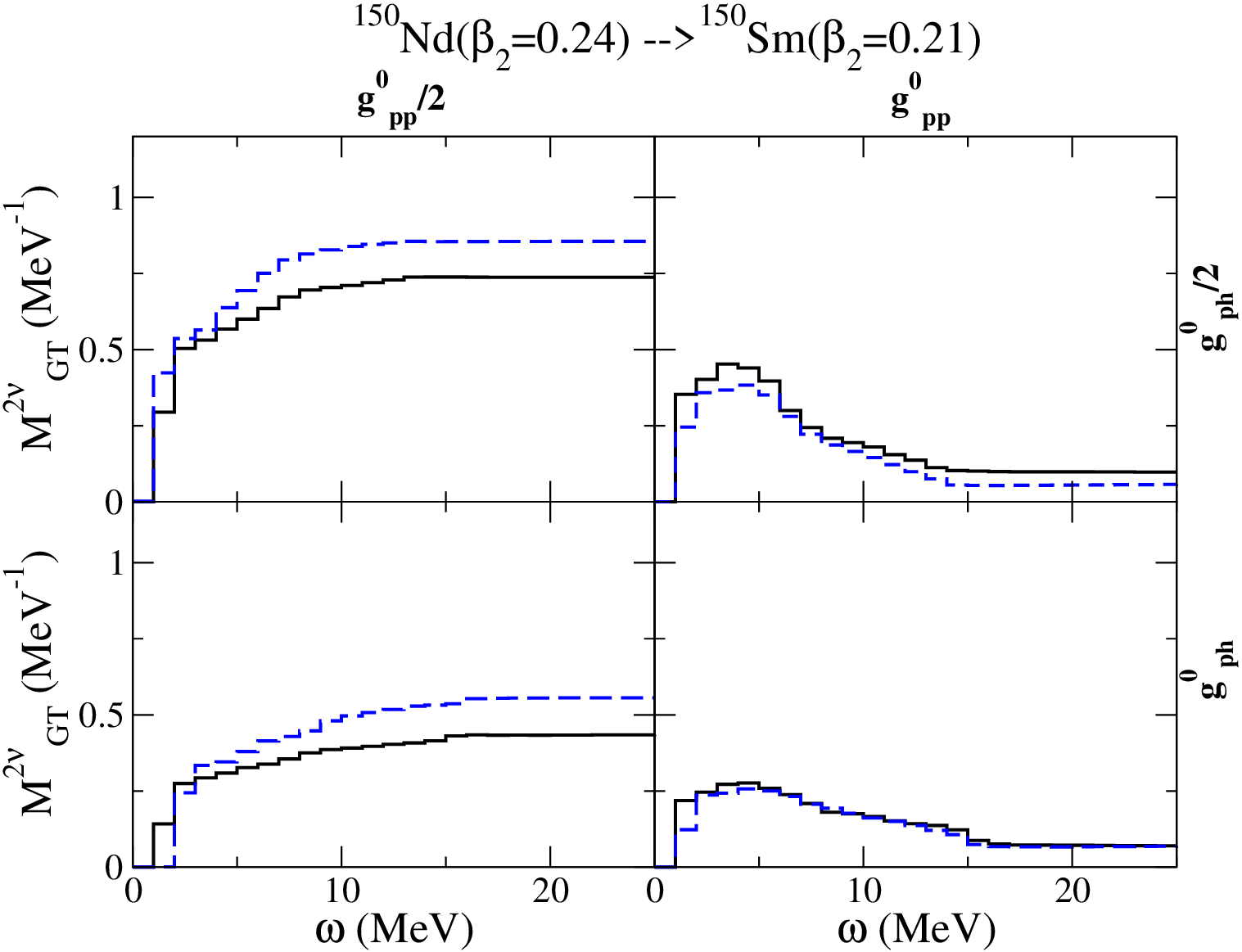}
\caption{(Color online) The same as in Fig.~\ref{ge}, but for $^{150}$Nd$\to^{150}$Sm $2\nu\beta\beta$ decay.}
\label{nd}
\end{figure}

To study the dependence of the shape of the calculated $2\nu\beta\beta$ running sums on different parameters of the residual interaction, we have calculated $M^{2\nu}_{GT}(\omega)$ not only for the values of $g_{ph}^0$ ($\chi^0$) and 
$g_{pp}^0$ ($\kappa^0$) listed in Table~\ref{tab1}, but also for the p-h and p-p strengths quenched by a factor of 2. The running sums $M^{2\nu}_{GT}(\omega)$ calculated with different combinations of the quenched and unquenched strengths are plotted in Figs.~\ref{ge} and \ref{nd}. Each panel in the figures shows four types of calculated running sums corresponding to 
different combinations of quenched and unquenched strength parameters 
(columns are labeled by values of $g_{ph}$, and rows by values of $g_{pp}$ in the calculation).

As shown in Figs.~\ref{ge},\ref{nd}, this is the p-p interaction parameter $g_{pp}$ ($\kappa$)
that is responsible for the qualitative change in the behavior of the calculated $2\nu\beta\beta$ running sums. 
For the quenched strengths $g_{pp}=g_{pp}^0/2$ ($\kappa=\kappa^0/2$), all calculations reveal a monotonically 
growing $M^{2\nu}_{GT}(\omega)$. The total calculated $M^{2\nu}_{GT}$ is, however, rather strongly overestimated for these quenched $g_{pp}$.
For the realistic, unquenched, $g_{pp}=g_{pp}^0$ ($\kappa=\kappa^0$), a substantial overshoot 
of the total $M^{2\nu}_{GT}$ at low excitation energies is compensated by a negative contribution 
coming from the region of the GTR at $\omega\approx$10 MeV. The absence of such a pronounced overshoot in the case of $^{150}$Nd$\to^{150}$Sm $2\nu\beta\beta$ beta decay calculated with  set I of the deformation parameters $\beta_2$ can again be attributed to a substantially smaller fitted value of $g_{pp}^0$ ($\kappa^0$) in this case, see Table~\ref{tab1}. In this case the difference in $\beta_2$ between the initial and the final nucleus is large, and this leads to a large overall suppression of the calculated $M^{2\nu}_{GT}$ by a small BCS overlap factor~\cite{Sal09}.

%\section {\bf Conclusions.}
We have  calculated the $2\nu\beta\beta$-decay running sums for $^{76}$Ge and $^{150}$Nd nuclei within the QRPA approach of Ref.~\cite{Sal09} which takes nuclear deformation into account and implements a realistic nucleon-nucleon residual interaction based on the Brueckner G matrix (for the Bonn CD force). We have studied 
%in this Brief Report 
the influence of different QRPA model parameters on the functional behavior of the running sums  within the QRPA approach.
It has been found that the parameter $g_{pp}$ renormalizing the G matrix in the particle-particle channel is responsible for a qualitative change in behavior of the running sums. 
Therefore, a different choice of the $g_{pp}$ parameter in different QRPA calculations may lead to a different, constructive or destructive, contribution to the total $2\nu\beta\beta$-decay matrix element coming from high-lying $1^+$ states.
For realistic values of $g_{pp}$ a significant negative contribution 
to the total $2\nu\beta\beta$-decay matrix element is found in the present calculation to come from the energy region of the giant GT resonance. This observation is in accord with other results~\cite{Rod05,Zhao90,Hor07,Nak96,Caur04}. The values of $g_{pp}$ adopted in Ref.~\cite{Mor08} are apparently smaller than those in the present calculations, and this may be a reason for the different functional behavior of the running sums calculated in Ref.~\cite{Mor08}.
Thus, a realistic value of $g_{pp}$ should be chosen thoroughly within a QRPA approach to draw a conclusion regarding the validity of the SSD.

%\acknowledgments
The authors acknowledge the support of the Deutsche Forschungsgemeinschaft under both SFB TR27 "Neutrinos and Beyond" and Graduiertenkolleg GRK683.
%, and the EU ILIAS project under the contract RII3-CT-2004-506222. 

\newpage

\end{document}